\documentclass[12pt,journal, onecolumn, draftcls]{IEEEtran}
\usepackage{amsmath}
\usepackage{graphicx}
\include{psfig}
\usepackage{epsfig}
\usepackage{psfig}
\usepackage{psfrag}
\usepackage{array}
\usepackage{amssymb}
\usepackage{color}
\begin{document}
\title{Sidelobe Control in Collaborative Beamforming via Node
Selection}

\author{Mohammed F. A. Ahmed,~\IEEEmembership{Student Member,~IEEE}
\\ and Sergiy A.~Vorobyov\thanks{This work is supported in parts
by the Natural Science and Engineering Research Council (NSERC) of
Canada and the Alberta Ingenuity Foundation, Alberta, Canada.

The authors are with the Department of Electrical and Computer
Engineering, University of Alberta, 9107-116 St., Edmonton,
Alberta, T6G~2V4 Canada. Emails: {\tt \{mfahmed,
vorobyov\}@ece.ualberta.ca}

{\bf Corresponding author:} Sergiy A.~Vorobyov, Dept. Elect. and
Comp. Eng., University of Alberta, 9107-116 St., Edmonton,
Alberta, T6G 2V4, Canada; Phone: +1 780 492 9702, Fax: +1 780 492
1811. Email: {\tt vorobyov@ece.ualberta.ca}.
},~\IEEEmembership{Senior Member,~IEEE}}
\maketitle

\vspace{-1.8cm}
\begin{abstract}
Collaborative beamforming (CB) is a power efficient method for
data communications in wireless sensor networks (WSNs) which aims
at increasing the transmission range in the network by radiating
the power from a cluster of sensor nodes in the directions of the
intended base station(s) or access point(s) (BSs/APs). The CB
average beampattern expresses a deterministic behavior and can be
used for characterizing/controling the transmission at intended
direction(s), since the mainlobe of the CB beampattern is
independent on the particular random node locations. However, the
CB for a cluster formed by a limited number of collaborative nodes
results in a sample beampattern with sidelobes that severely
depend on the particular node locations. High level sidelobes can
cause unacceptable interference when they occur at directions of
unintended BSs/APs. Therefore, sidelobe control in CB has a
potential to increase the network capacity and wireless channel
availability by decreasing the interference. Traditional sidelobe
control techniques are proposed for centralized antenna arrays
and, therefore, are not suitable for WSNs. In this paper, we show
that distributed, scalable, and low-complexity sidelobe control
techniques suitable for CB in WSNs can be developed based on node
selection technique which make use of the randomness of the node
locations. A node selection algorithm with low-rate feedback is
developed to search over different node combinations. The
performance of the proposed algorithm is analyzed in terms of the
average number of trials required to select the collaborative
nodes and the resulting interference. Our simulation results
approve the theoretical analysis and show that the interference is
significantly reduced when node selection is used with CB.
\end{abstract}

\vspace{-0.5cm}
\begin{IEEEkeywords}
Collaborative beamforming, sidelobe control, wireless sensor
networks, node selection.
\end{IEEEkeywords}

\section{INTRODUCTION}
\label{sec:intro} Wireless sensor networks (WSNs) have become
practical technology due to the production of low cost, low-power,
and small size sensors. Different applications such as habitat and
climate monitoring, detection of human/vehicular intrusion, and
etc. are increasingly employing WSNs \cite{Culler2004}. Such
applications require sensor nodes to be deployed over a remote
area to collect data from the surrounding environment and
communicate it to far base stations or access points (BSs/APs). As
a result, the challenges faced in the WSN applications are quite
different from that of considered in the applications of the
traditional wireless ad-hoc networks \cite{Akyildiz2002}. These
differences can be summarized as follows.
\begin{itemize}
\item[(i)] Typical WSN is densely deployed and may consist of
thousands of sensor nodes.
\item[(ii)] The network geometry changes all the
time due to failure of sensor nodes or deployment of new sensor
nodes.
\item[(iii)] Sensor nodes are battery-powered and the battery often
cannot be replaced. Thus, the sensor node life time is limited by
the battery lifetime.
\item[(iv)] Sensor nodes have simple hardware with limited
computational capabilities and small memory in order to keep the
production cost of the sensor node reasonable.
\item[(v)] Sensor nodes can fail easily. Thus, it is desired that
the WSN performance does not depend on individual sensor nodes.
\item[(vi)] Data and traffic models in WSNs depend on the
application, and usually the data is redundant, while the traffic
has low-rate burst nature.
\item[(vii)] Sensor nodes in WSNs are usually deployed at the
ground level and have no mobility. Thus, the channel path loss for
individual node is high and the channel variations are slow.
\end{itemize}
Practical communication schemes for WSNs should
overcome the problem of limited transmission range of individual
sensor nodes, while being distributed and scalable. Moreover, for
designing such communication schemes, power consumption and
implementation complexity issues have to be taken into account as
the most significant design constraints for WSNs.

To address the aforementioned issues, the inherent high density
deployment of sensor nodes has been used to introduce
collaborative beamforming (CB) for the uplink
communication to a BS/AP \cite{Ochiai2005}, \cite{Ahmed2009a}. CB
extends the transmission range of individual sensor nodes by using
a cluster of sensor nodes in a power-efficient way. Particularly,
sensor nodes from a cluster of nodes act collaboratively as
distributed antenna array to form a beam toward the
direction(s) of the intended BS(s)/AP(s). Given that each sensor
node is equipped with a single omnidirectional antenna and
operates in half-duplex mode, CB is performed in two stages. In
the first stage, the data from source node(s) in a cluster is
shared with all other collaborative nodes, while in the second
stage, this data is transmitted by all sensor nodes simultaneously
and coherently. In the latter stage, sensor nodes adjust the
initial phase of their carriers so that the individual signals
from different sensor nodes arrive in phase and constructively add
at the intended BS/AP. In this way, CB is able to increase the
area coverage of WSNs and, therefore, can be also viewed as an
alternative scheme to the multi-hop relay communications. However,
as compared to the multi-hop relay communications, CB brings the
following advantages.
\begin{itemize}
\item[(i)] For CB, there is no dependency of communication quality
on individual nodes. Thus, the communication link is more
reliable.
\item[(ii)] CB distributes the power consumptions over large number
of sensor nodes and balances the lifetimes of individual nodes
\cite{Han2007}.
\item[(iii)] CB enables to create a direct single-hop uplink to the
intended BS(s)/AP(s). Thus, it reduces the communication delay and data
overhead.
\item[(iv)] CB achieves higher connectivity than that of
omnidirectional transmission with the same transmit power
\cite{Kiese}.
\end{itemize}

In order to implement CB, the following issues related to the the
distributed nature of the WSNs have been addressed. Distributed
schemes for estimating the initial phases of the local node
oscillators in WSNs have been introduced in
\cite{Mudumbai2006}--\cite{Brown2008}. These schemes allow to
achieve phase synchronization among all collaborative nodes in a
cluster of WSN. Moreover, to minimize the time required for
multiple sources to share the data among all sensor nodes in a
cluster, a medium access control-physical (MAC-PHY) CB scheme
which is based on the medium random access has been proposed in
\cite{Dong2008}.

Although, the above mentioned implementation issues for CB have
been positively addressed, one more concern is that the random
sensor node locations result in a random beampattern which depends
on the actual locations. The effect of the spatial sensor node
distribution on the directivity of the CB beampattern has been
studied in \cite{Ochiai2005} (see also \cite{Lo1964}) for the case
of uniform sensor node distribution and in \cite{Ahmed2009a},\cite{Ahmed2008a},
\cite{Ahmed2008b} for the case of Gaussian sensor node distribution. Although
it has been shown for both aforementioned node distributions that
the CB sample beampattern has a deterministic mainlobe which is
independent on the random sensor node locations, the sidelobes of the
CB sample beampattern are totaly random and can be described only
in statistical terms \cite{Ochiai2005},
\cite{Ahmed2009a,Ahmed2008b}. In addition, the aforementioned
multiple access scheme of \cite{Dong2008}, which minimizes the
time required for sharing the multiple source data, results in
higher sidelobes even for the average CB beampattern. All these
can lead to high interference levels at the directions of
unintended BSs/APs. Therefore, the sidelobe control problem arises
in the context of WSNs. Indeed, achieving a {\it sample} CB
beampattern with lower sidelobe interference at unintended BSs/APs
has the potential to increase the WSN throughput
\cite{Ramanathan}.

Due to the inherent distributed nature of WSNs, the sidelobe control
should be achieved with minimum data overhead and knowledge of the
channel information. Unfortunately, traditional sidelobe control
techniques \cite{Liu2003,Hughes1996} cannot be applied in
the context of WSNs due to their unacceptably high complexity and
the requirement of centralized processing. Indeed, to apply the
centralized beamforming weight design in the WSNs, a node or BS/AP
has to collect the location and channel information from all
sensor nodes. It can significantly increase the overhead in the
network and nullify the above mentioned advantages of CB. Note that
for the same reasons the recently developed network beamforming
techniques \cite{Jing2008,Havary-Nassab2008} are
restricted to the applications in the relay networks only and it
is impractical to apply them for WSNs.

In this paper\footnote{Some preliminary results have been also
reported in \cite{Ahmed2009b}.}, we develop a technique for
sidelobe control in CB for WSNs which is based on sensor node
selection. Such technique makes use of the randomness of node
locations and is distributed and scalable as well as it has low
data overhead. Moreover, as compared to the optimal beamforming
weights assignment, our sidelobe control technique which is based
on the phase synchronization and simple node selection is more
robust to the channel/phase errors. In addition, it helps to
balance the life times of all sensor nodes since the corresponding
beamforming weights have the same magnitude for all nodes. For the
sidelobe control technique, a node selection algorithm with
low-rate feedback is developed to search over different node
combinations. The performance of the proposed algorithm is
analyzed in terms of the average number of trials required to
select the collaborative nodes and the distribution of the
resulting interference.

The paper is organized as follows. System and signal models are
introduced in Section~II. A new sidelobe control technique for CB
in WSNs is developed in Section~III where the node selection
algorithm is also summarized. The performance characteristics of
the proposed node selection algorithm such as the average number
of trials required to select the collaborative nodes and the
resulting interference are studied in Section~IV. Section~V
reports our simulation results and is followed by conclusions in
Section~VI. Proofs of some results in the paper are summarized in
Appendixes.

\section{SYSTEM AND SIGNAL MODEL}

\subsection{System Model}
\begin{figure}[t]
\centering
\includegraphics[width=.8\textwidth]{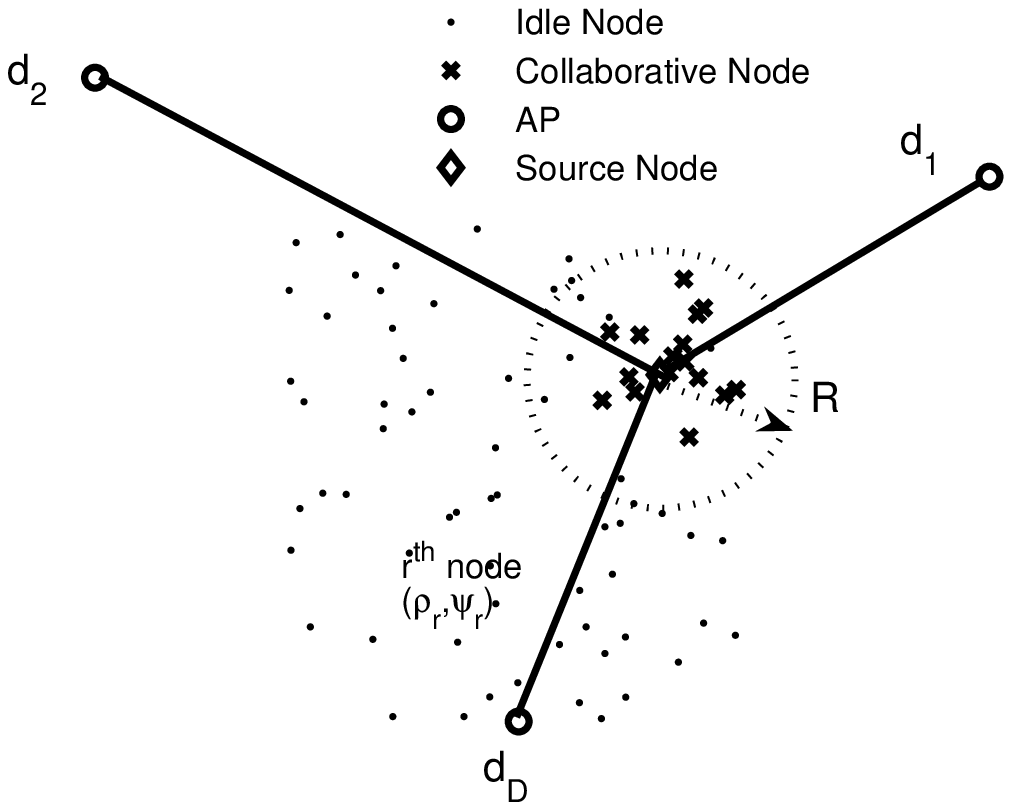}
\caption{WSN model with multiple BSs/APs.}
\label{Fig:Journal2_fig1}
\end{figure}

We consider a WSN with nodes randomly placed over a plane as shown
in Fig.~\ref{Fig:Journal2_fig1}. Multiple BSs/APs, denoted as
${\cal D} = \{ d_0, d_1, d_2, \dots, d_D \}$, are located outside
and far apart from the coverage area of each individual node at
directions $\varphi_0, \varphi_1, \varphi_2, \dots, \varphi_D $,
respectively. Uplink transmission is a burst traffic for which
the nodes are idle most of the time and have sudden transmissions.
Thus, we adopt a time-slotted scheme where nodes are allowed to
transmit at the beginning of each time slot. The downlink
transmissions are mostly the control data broadcasted over a
separate error-free control channels. The BSs/APs can use high
power transmission and, therefore, the downlink is less
challenging and can be organized as direct transmission.

We assume that due to the limited power of individual nodes,
direct transmission to the BSs/APs is not feasible and sensor
nodes have to employ CB for the uplink transmission. The distance
between nodes in one cluster of WSN is small so that the power
consumed for communication among nodes in the cluster can be
neglected. Each sensor node is equipped with a single antenna used for
both transmission and reception. To identify different nodes and
BSs/APs, each node or BS/AP to has a unique
identification (ID) sequence that is included in each
transmission.

At each time slot, a set ${\cal S} = \{ s_0, s_1, s_2, \dots, s_S
\}$ of source nodes is active. Moreover, only $K$
source--destination pairs are allowed to communicate. Here $K =
\min \{ {\rm card} ( \cal S ) , {\rm card} ( \cal D ) \}$ where ${\rm
card} (\cdot )$ denotes the cardinality of a set, and the $k$th
source--destination pair is denoted as $s_k$--$d_k$. For source node
$s_k$, the coverage area is, ideally, a circle with a radius which
depends on the power allocated for the node-to-node communication.

Let ${\cal M}^k$ be a set of nodes in the coverage area of the
node $s_k$. Let us, therefore, select the source node $s_k$ as a
local origin of the coordinate system used to mark the spatial
locations of the nodes in the coverage area of $s_k$, i.e. ${\cal M}^k$. The
$r$th collaborative node, denoted as $c_r$, $r \in {\cal M}^k$,
has a polar coordinates $\left( \rho_r , \psi_r \right)$. The
Euclidean distance between the collaborative node $c_r$ and a
point $(A, \phi)$ in the same plane is defined as
\begin{equation} \label{dist}
d_r (\phi) \triangleq \sqrt{{A}^2 + \rho_r^2 - 2 \rho_r {A}
\cos(\phi -\psi_k)} \approx A - \rho_r \cos(\phi-\psi_r)
\end{equation}
where $A \gg r_k$ in the far-field region.

The array factor for the set of sensor nodes ${\cal M}^k$ in a
plane can be defined as
\begin{equation}
AF_{k} (\phi) =  \sum_{r \in {\cal M}^k} \sqrt{P_r} e^{j
\theta_r^{k}} e^{ -j \theta_r(\phi)} \label{Eq:AFdef}
\end{equation}
where $P_r$ is the transmission power assigned to each node, $\theta_r^{k}$ is the initial
phase of the $r$th sensor carrier frequency, $\theta_r (\phi)
= \frac{2\pi} {\lambda} d_r(\phi)$ is the phase delay due to
propagation at the point $(A, \phi)$, and $\lambda$ is the wavelength. Then the far-field beampattern corresponding to the
set of sensor nodes ${\cal M}^k$ can be found as
\begin{equation}
BP_{k}(\phi) \triangleq \left| AF_{k}(\phi) \right|^2 = \left|
\sum_{r \in {\cal M}^k} \sqrt{P_r} e^{j \theta_r^{k}}
e^{ -j \theta_r(\phi)} \right|^2 \label{Eq:patterndefine}
\end{equation}
where $| \cdot |^2$ denotes the magnitude of a complex number.

It is assumed that the symbol duration is very short as compared
to the channel coherent time. Therefore, the channel variations
can be considered to be quasi-static during one symbol
transmission. Since sensor nodes are located at the ground level,
the large-scale fading is the dominant factor for the channels
between collaborative nodes in ${\cal M}^k$ and BSs/APs. Then the
channel coefficient for $r$th collaborative node which serves $k$th
source--destination pair can be modeled as
\begin{equation} \label{channelgain}
h_{rk} = a_{rk} b_{rk}
\end{equation}
where $b_{rk}$ is the attenuation/path loss factor in the channel
coefficient due to propagation distance and $a_{rk}$ is a
lognormal distributed random variable which represents the
flactuation/shadowing effect in the channel coefficient, i.e.,
$a_{rk} \sim \exp \{ {\cal N}(m, \sigma^2) \}$. Here $m$ and
$\sigma^2$ denote respectively the mean and variance, of the
corresponding Gaussian distribution. Then the mean and variance of
the lognormally distributed $a_{rk}$ can be found as
\begin{eqnarray}
m_{a_{rk}} \!\!&=&\!\! E\left\{ a_{rk} \right\} = e^{ m +
{\frac{\sigma^2}{2}}} \\
\sigma_{a_{rk}}^2 \!\!&=&\!\!
E\left\{a_{rk}^2\right\} = \left( e^{\sigma^2} - 1 \right) \left(
e^{2m + \sigma^2} \right)
\end{eqnarray}
where $E\{ \cdot \}$ stands for the statistical expectation. The
attenuation/path loss depends on the distance between $c_r$ and
$d_k$ and the
path loss exponent. 
Assuming that all nodes in ${\cal M}^k$ are close to each other,
the pass losses from the nodes in ${\cal M}^k$ to the BS/AP are
equal to each other, i.e., $b_{rk} = b_k$, $r \in {\cal M}^k $
\cite{Lin200x}. Moreover, since all BSs/APs are located far apart
from the cluster of collaborative nodes, the network can be viewed
as homogeneous and the attenuation effects of different paths can
be assumed approximately equal to each other, i.e., $b_k = b$
\cite{Dong2009}. Note that even if the attenuation effects for
different BSs/APs are different, they can be compensated by
adjusting the gains of the corresponding receivers or the
power/number of the corresponding collaborative nodes participating
in CB. 

\subsection{CB and Corresponding Signal Model}
Consider a two--step transmission which consists of the
information sharing and the actual CB steps. Information sharing
aims at broadcasting the data from one source node to all other
nodes in its coverage area. Specifically, in this step, the source
node $s_k$ broadcasts the data symbol $z_k$ to all nodes in its
coverage area ${\cal M}^k$, where the data symbol $z_k \in
{\mathbb{C}}$ belongs to a codebook of zero mean, unit power, and
independent symbols, i.e., $E \left\{ z_k \right\} = 0$, $\left|
z_k^2 \right| = 1$, and $E\left\{z_k z_n \right\} = 0$ for $n \ne
k$.

In the case of multiple source nodes sharing their own data with
other nodes in their corresponding collaborative sets of nodes, a
collision can occur. To avoid the collision, orthogonal channels
in frequency, time, or code can be used. However, such collision
avoidance causes resource loss and can lead to the network
throughput reduction, especially if the number of source nodes
sharing the data is large. Therefore, collision resolution schemes
can be used alternatively (see for example \cite{Lin2005},
\cite{Yang2008}). In this case, the information sharing takes only
one time slot in a random access fashion.
Finally, we assume that the power used for broadcasting the data
by the source node is high enough so that each collaborative node
$c_r$ can successfully decode the received symbol from the source
node $s_k$.

During the CB step, each collaborative node $c_r, r \in {\cal
M}^k$, is first synchronized with the initial phase $\theta_r ^k =
-\frac{2\pi}{\lambda} \rho_r \cos (\varphi_k - \psi_r)$ using the
knowledge of the node locations (see the closed-loop scenario in
\cite{Ochiai2005}). Alternatively the synchronization can be
performed without any knowledge of the node locations (see
\cite{Mudumbai2006}--\cite{Brown2008}). For example, the
synchronization algorithm of \cite{Mudumbai2006} uses a simple
1-bit feedback iterations, while the methods of \cite{Wang2008}
and \cite{Brown2008} are based on the time-slotted round-trip
carrier synchronization approach.

After synchronization, all collaborative nodes transmit the signal
coherently
\begin{equation}
t_r = z_k e^{j\theta_r^k}, \quad r \in {\cal M}^k.
\end{equation}
Then the received signal at angle $\phi$ can be given as
\begin{eqnarray} \label{ModelMain}
g( \phi ) = \sum_k z_k \sum_{r \in {\cal M}^k} \sqrt{P_r} a_{rk}
e^{j\theta_r^k} e^{-j\theta_r(\phi)}  + w
\end{eqnarray}
where $w \sim {\cal CN} ( 0, \sigma_w^2 )$ is the additive white
Gaussian noise (AWGN) at the direction $\phi$. Note that the white
noise is the same at all angles and, therefore, it disturbs the CB
beampattern \eqref{Eq:patterndefine} equally in all directions.
The received noise power $\sigma_w^2$ at BSs/APs can be measured
in the absence of data transmission and, therefore, is assumed to
be known at each BS/AP.

The received signal at the intended BS/AP $d_{k^{*}}$ can be
written as
\begin{eqnarray}
g_{k^{*}} \triangleq g( \varphi_{k^{*}} ) &=& z_{k^{*}}  \sum_{r
\in {\cal M}^{k^{*}}} \sqrt{P_r} a_{r{k^{*}}} + \sum_{k \ne k^{*}}
z_k \sum_{r \in {\cal M}^k} \sqrt{P_r} a_{r{k^{*}}}
e^{-j(\theta_r^{k ^{*}}-\theta_r^{k})} +
w \nonumber \\
&=& z_{k^{*}} \sum_{r \in {\cal M}^{k^{*}}} \sqrt{P_r}
a_{r{k^{*}}} + \sum_{k \ne k^{*}} z_k \sum_{r \in {\cal M}^k}
\sqrt{P_r} a_{r{k^{*}}} \left( x_r^{(k^{*},k)} - j y_r^{(k^{*},k)}
\right) + w \label{Eq:y1}
\end{eqnarray}
where $x_r^{(k^{*},k)} = {\cal R} \left\{ e^{-j( \theta_r^{k ^{*}}
- \theta_r^{k })} \right\}$, $y_r^{(k^{*},k)}={\cal I} \left\{
e^{-j ( \theta_r^{k ^{*}} - \theta_r^{k})} \right\}$ and ${\cal
R}\left\{ \cdot \right\}$ and ${\cal I}\left\{ \cdot \right\}$
represent the real and the imaginary parts of a complex number,
respectively. Note that $x_r^{(k^{*},k)}$ and $y_r^{(k^{*},k)}$
are random variables. It can be further shown (see Appendix A)
that $u \in \left\{ x_r^{(k^{*},k)}, y_r^{(k^{*},k)} \right\}$ has
mean $m_{x_r^{(k^{*},k)}} = m_{y_r^{(k^{*},k)}} = m_u = E \left\{
u \right \} = 0$ and variance $\sigma_{x_r^{(k^{*},k)}}^2 =
\sigma_{y_r^{(k^{*},k)}}^2 = \sigma_u^2 = E \left\{ u^2 \right\} =
0.5$. The first term in \eqref{Eq:y1} is the signal received at
the the BS/AP $d_{k^{*}}$ from the desired set of collaborative
nodes ${\cal M}^{k^{*}}$, while the second term represents the
interference caused by other sets of nodes ${\cal M}^k$, $k \ne
k^{*}$ where ${\cal M}^k \cap {\cal M}^n = \varnothing$, $k \ne
n$.
Assumed that each node in the network utilizes the same amount of
power for each CB transmission, i.e., $P_r=P$, the received signal
\eqref{Eq:y1} at the BS/AP $d_{k^{*}}$ can be rewritten as

\begin{eqnarray}
g_{k^{*}} = \sqrt{P} \ \ z_{k^{*}} \sum_{r \in {\cal M}^{k^{*}}}
a_{r{k^{*}}} + \sqrt{P} \sum_{k \ne k^{*}} z_k \sum_{r \in {\cal
M}^k} a_{r{k^{*}}} \left( x_r^{(k^{*},k)} - j y_r^{(k^{*},k) }
\right) + w. \label{Eq:y}
\end{eqnarray}

%
%

\section{Sidelobe Control via Node Selection}
In dense WSNs, each source node is surrounded by many candidate
collaborative nodes in its coverage area. Although it has been
shown earlier that the sidelobe levels of the average beampattern
decrease inverse proportionally and uniformly over all directions
with the increase of the number of collaborative nodes
\cite{Ochiai2005}, \cite{Ahmed2009a}, the sidelobe levels at
particular directions of interest (unintended BSs/APs) in the
sample beampattern are totaly random and can be unacceptably high
if the number of collaborative nodes is not very large. At the
same time, the randomness of the node locations provides
additional degrees of freedom for controlling the beampattern
sidelobes. Indeed, the sidelobes corresponding to different sets
of collaborative nodes are different \cite{Ochiai2005},
\cite{Ahmed2009a}.

The problem of high sidelobe levels at specific directions for the
average beampattern has been briefly discussed in
\cite{Zarifi2009b}. It is suggested there to use only the sensor
nodes placed in multiple concentric rings instead of using all
nodes in the disk of the coverage area. However, a narrower ring
with larger radiuses results in the average beampattern with
smaller mainlobe width and leads to larger sidelobe peak levels at
other uncontrolled directions than the conventional CB. Moreover,
only the average beampattern behavior is considered in
\cite{Zarifi2009b}, while it is the sample beampattern behavior
that is of real importance for the sidelobe control in WSNs.

Exploiting the randomness of the node locations in WSNs, we
introduce and study in this section a node selection algorithm for
sidelobe control of the CB sample beampattern as a method for
interference reduction. A sidelobe control approach based on node
selection is suitable for WSN applications because it allows to
avoid complex central beamforming weight design and corresponding
additional communications. 

It is required to select a subset of collaborative nodes from the
candidate nodes in the coverage area of the source node. Let
${\cal N}^k$ be a set of collaborative nodes to be selected from
${\cal M}^k$, i.e., ${\cal N}^k \subset {\cal M}^k$, to beamform
data symbols to $d_k$. Note that in accordance with
\cite{Ochiai2005} and \cite{Ahmed2009a}, the mainlobe of the
beampattern is stable and does not change for different subsets of
${\cal M}^k$ as long as the the size of the coverage area does not
change and the WSN is sufficiently dense, i.e., each cluster of
the WSN consists of a sufficiently large number of sensor nodes.
Also note that the set ${\cal N}^k$ can be updated any time when
the channel conditions or network configuration change.
Alternatively, it can be updated periodically to balance power
consumptions among nodes. The meaningful objective for node
selection is to achieve a beampattern with low level sidelobes
toward the unintended BS/AP directions.

Toward this end, we develop a low-complexity distributed node
selection algorithm, which guarantees that the sidelobe levels
toward unintended direction(s) are below a certain prescribed
value(s) as long as the WSN is sufficiently dense. An algorithm
utilizes only the knowledge of the received interference power to
noise ratio (INR), denoted as $\eta$, at the unintended
destinations and requires only low-rate (essentiality, one-bit)
feedback from the unintended BSs/APs at each trial. Although our
node selection strategy does not guarantee the optimum result of
centralized beamforming strategies (which require global CSI and,
therefore, a very significant data overhead in the network), it
has the following practically important advantages for applying in
the WSNs context.
\begin{itemize}
\item[(i)]~It is very simple computationally and can be run in cheap
sensor nodes without adding any computations.
\item[(ii)]~It has a distributed nature and, therefore, uses minimum
control feedback from the unintended BSs/APs.
\end{itemize}

In the following, we first describe the communication protocol and
then give the details of the node selection algorithm.

\subsection{Communication Protocol}
Data transmission is organized in the following steps.
\begin{itemize}
\item[{\it Step 1:}] At the beginning of each time-slot, the
source node $s_k$ listens to the control channels from BSs/APs
and checks for an available BS/AP.
\item[{\it Step 2:}] The source node $s_k$ broadcasts its
ID and the ID of the available BS/AP $d_k$ to the nodes in
its coverage area ${\cal M}^k$. Note that all transmissions
from the source node $s_k$ to the intended destination
$d_k$ will be achieved through the CB using the nodes in the set of
collaborative nodes ${\cal N}^k \subset {\cal M}^k$.
\item[{\it Step 3:}] The source node $s_k$ attempts to
transmit its ID to the target BS/AP. If a collision occurs,
i.e., if the target BS/AP receives transmissions form more than one
collaborative sets of sensor nodes, the BS/AP approves only one
set of nodes for data transmission.
\item[{\it Step 4:}] At the beginning of each following time-slot,
the BS/AP broadcasts through the control channels the selected
source ID in addition to one bit of information which indicates
that the BS/AP is busy. In this case, other source nodes are not
allowed to transmit data to the same BS/AP.
\item[{\it Step 5:}] Finally, only the predetermined subset of
collaborative nodes ${\cal N}^k$ assigned to the pair
$s_k$--$d_k$ continues to receive data, while other nodes
go back to idle mode, that is, the actual data transmission takes a
place for the pair $s_k$--$d_k$.
\end{itemize}

\subsection{Node Selection Algorithm}
A set of collaborative nodes ${\cal N}^k \subset {\cal M}^{k}$ is
assigned to each source--distention pair $s_k$--$d_k$. To select
such a collaborative set, the nodes can be tested one by one or a
group of nodes by a group of nodes. The latter is, however,
preferable since it can significantly reduce the data overhead in
the system. Indeed, while testing one node or a group of nodes, we
need to check if the corresponding CB beampattern sidelobe level
reduces in the unintended direction(s) and then send the
`approve/reject' bit per one node in the fist case, or per a group
of nodes in the second case. Therefore, if every group of nodes
consists of a larger number of sensor nodes, less `approve/reject'
bits has to
be sent in the system in total. 

Consider the source node $s_{k^{*}}$, let the number of nodes in
its coverage area is $M$, the number of collaborative nodes needed
to be selected is $N \le M$, and the size of one group of nodes to
be tested in each trial is $L \le N$. Using the selection
principle highlighted above, the selection process can be
organized in the following two steps.
\begin{itemize}
\item[{\it Step 1:}] {\bf Selection.} Source node $s_{k^{*}}$
initiates the node selection by broadcasting the {\it select
message} to the nodes in its coverage area, namely the set ${\cal
M}^{k^{*}}$, and randomly selects a subset ${\cal L}^{k^{*}}$ of
$L$ candidate nodes from ${\cal M}^{k^{*}}$.

The nodes can be assigned to the set ${\cal L}^{k^{*}}$ by using
any of the following two methods. The first one is a centralized
method in which the source node $s_{k^{*}}$ is totally responsible
for the node assignment. According to this method, every source
node maintains a table of IDs of all candidate nodes in its
coverage area and randomly assigns nodes to the set ${\cal
L}^{k^{*}}$. The source node $s_{k^{*}}$ then broadcasts the IDs
of the nodes assigned to the set ${\cal L}^{k^{*}}$ to inform them
that they are selected for the test. The disadvantage of this
method is that it is suitable only for small WSNs where each
source node can keep records of all other nodes in its coverage
area. As a result, this method typically requires large data
exchange between the source and the candidate nodes.

Alternatively, in the second method, node assignment task is
distributed among the source and collaborative nodes. In particular,
if collaborative nodes receive the {\it select message}, each node
starts a random delay using an internal timer. After the random
delay, the candidate node responds by the {\it offer
message} which contains the ID of this node. Then the source node
responds by the {\it approval message} which requires only 1 bit
of feedback. If a collision occurs and two collaborative nodes
transmit the {\it offer message} at the same time,
the source node responds by the {\it approval message} with a
different bit value and the timers in both nodes start over a new
random delay. The process repeats and the source node $s_{k^{*}}$
keeps sending the {\it select message} until $L$ candidate nodes
are assigned and the set ${\cal L}^{k^{*}}$ is constructed.

\item[{\it Step 2:}] {\bf Test.} The set ${\cal L}^{k^{*}}$ transmits
the {\it test message} that contains the intended BS/AP ID to the
intended destination $d_{k^{*}}$ using CB. While the intended
destination $d_{k^{*}}$ receives a predetermined {\it signal}
power level\footnote{The power level at the intended destination
depends on the number of collaborative sensor nodes and the power
of each of them.}, the {\it interference} power levels at the
unintended destination(s) $d_k$, $k \ne k^{*}$ are random because
of the random sidelobes of the CB beampattern. At this stage, all
unintended BSs/APs with different IDs measure the received INR
$\eta$. If $\eta$ is higher than a predetermined threshold value
$\eta_{\text{thr}}$, the {\it reject message} is sent back to the
candidate set ${\cal L}^{k^{*}}\!\!\!$. In this case, the nodes in
the candidate set ${\cal L}^{k^{*}}$ are all returned to the set
of nodes ${\cal M}^{k^{*}}$ and can be used in future trials. If
no {\it reject message} is received from any of the unintended
BSs/APs after a predetermined time, then the candidate set ${\cal
L}^{k^{*}}$
is approved and each node from the candidate set ${\cal
L}^{k^{*}}$ stores the IDs of the source node $s_{k^{*}}$ and the
destination $d_{k^{*}}$. Then the collaborative nodes assigned to
serve the pair $s_{k^{*}}$--$d_{k^{*}}$ do not participate in
future trials. In this way, we can avoid an overlap between sets
of nodes serving different BSs/APs.

In order to select $N$ collaborative nodes, the Selection and Test
steps are repeated until $N/L$ candidate sets ${\cal
L}_{l}^{k^{*}}$, $l = 1, 2, \dots, N/L$, are approved.\footnote{It
is assumed for simplicity that $N/L$ is an integer number. If
$N/L$ is not integer, it is still easy to adjust the size of the
candidate set ${\cal L}^{k^{*}}$ in the last trial of the
algorithm only. For example, the size of the last candidate set
${\cal L}^{k^{*}}$ can be chosen to be equal to the reminder of
$N/L$.} Then the so obtained set of approved collaborative nodes
is ${\cal N}^{k^{*}} = \bigcup_{l} {\cal L}_{l}^{k^{*}}$.

Once $N$ nodes are selected, i.e., ${\cal N}^{k^{*}}$ is
constructed, the source node $s_{k^{*}}$ broadcasts the {\it end
message} and no more candidate sets ${\cal L}^{k^{*}}$ is
constructed.
\end{itemize}
The pseudocode of the node selection algorithm is given in
Table~\ref{Tab:pseudocode}.
\begin{table}
\begin{center}
\begin{tabular}{|l|}
\hline
{\bf Node selection algorithm}\\
\hline
{\bf Initial values:}\\
$N$ and $L$ are predetermined at the Source Node $s_{k^{*}}$. \\
$\eta_{\text{thr}}$ is predetermined at the unintended
Destinations $d_{k}$,
$k = \{ 0, 1, \dots, D \}$. \\
\\
1: At $s_{k^{*}}$: ($\text{Counter }{l} \leftarrow 1$). \\
2: {\bf If} ($\text{Counter }{l} < \frac{N}{L}$), \\
3: \ \ \ \ \ \ \ {\bf Then:} \{ $s_{k^{*}}$ broadcasts the
{\it select message}. \\
4: \ \ \ \ \ \ \ \ \ \ \ \ \ \ \ \ \ \ A candidate set
${\cal L}_{l}^{k^{*}}$ is constructed. \\
5: \ \ \ \ \ \ \ \ \ \ \ \ \ \ \ \ \ \ Using CB, the nodes in
${\cal L}_{l}^{k^{*}}$ transmit the {\it test message}.\} \\
6: \ \ \ \ \ \ \ \ {\bf Otherwise:} \{Go to 12.\} \\
7: At any $d_{k}$, $k \ne k^{*}$: {\bf If} ( The received
INR $\eta > \eta_{\text{thr}}$),\\
8: \ \ \ \ \ \ \ \ {\bf Then} \{ $d_{k}$ sends the {\it reject
message} to ${\cal L}_{l}^{k^{*}}$.\} \\
9: \ \ \ \ \ \ \ \ {\bf Else} \ \ \{ No {\it reject
message} is received. \\
10:  \ \ \ \ \ \ \ \ \ \ \ \ \ \ \ \  ${\cal L}_{l}^{k^{*}}$
is approved and the corresponding nodes store the IDs of
$s_{k^{*}}$ and $d_{k^{*}}$. \\
11: \ \ \ \ \ \ \ \ \ \ \ \ \ \ \ \ At $s_{k^{*}}$:
($\text{Counter }{l} \leftarrow \text{Counter } {l + 1} $).
{\bf Go to} 2.\}\\
12: $s_{k^{*}}$ broadcasts the {\it end message}.\\
\hline
\end{tabular}
\label{Tab:pseudocode}
\caption{Table~1:~Node selection algorithm for CB sidelobe control.}
\end{center}
\end{table}

\section{Performance Analysis}
In this section, we analyze the proposed node selection algorithm
in terms of (i)~the average number of trials required for
selecting a set of collaborative nodes which guarantee low
interference level for unintended BSs/APs and (ii)~the
complementary cumulative distribution function (CCDF) of the
INR $\eta$. The first characteristic allows to estimate the
average run time of the algorithm, while the second characteristic
is needed to estimate the achievable interference levels versus
the corresponding interference threshold values.

\subsection{The Average Number of Trials}
Note that the signal-to-noise ratio (SNR), denoted as $\gamma$,
received through the link $s_{k^{*}}$--$d_{k^{*}}$ at the intended
BS/AP should be above a certain level which guarantees the correct
detection with high probability. This SNR at the intended BS/AP
must be guaranteed regardless of the number of collaborative nodes
participating in the CB. Therefore, we assume in our analysis that
the total transmit power budget for each tested candidate set of
nodes is kept the same in each trial of CB transmission. In
particular, if the SNR at the intended BS/AP is required to be $10
\log_{10}(\gamma)$~dB, then in the selection process, the power
per one sensor node in ${\cal L}_{l}^{k^{*}}$ has to be set as $P
= \sigma_w^2 \gamma /L \le P^{\rm max}$, where $P^{\rm max}$ is
the maximum available power at the node. Also note that the power
consumed for running the node selection algorithm is, in fact,
proportional to the average number of trials in the algorithm.
Therefore, it is preferable to construct a set of collaborative
nodes with less number of trials.

In order to derive the average number of trials for the node
selection algorithm, we, first, need to find the probability that
a candidate set of nodes ${\cal L}_{l}^{k^{*}}$ is approved as
part of the set of collaborative nodes ${\cal N}^{k^{*}}$. This
probability is the same as the probability that the set ${\cal
L}_{l}^{k^{*}}$ generates an acceptable interference at the
unintended BSs/APs. Since we assumed that only one set of
collaborative nodes ${\cal N}^{k^{*}}$ is constructed at a time,
there is no interference present from other collaborative sets.
Therefore, the interference power received at the unintended BS/AP
$d_{k}$ from the tested candidate set of nodes ${\cal
L}_{l}^{k^{*}}$ which targets the intended BS/AP $d_{k^{*}}$ can
be written as
\begin{eqnarray}
I \left(\varphi_k\left| {\cal L}_{l}^{k^{*}} \right.\right) =
\sqrt{\frac{\sigma_w^2 \gamma}{L}} z_{k^*} \!\! \sum_{r \in {\cal
L}_{l}^{k^{*}}} a_{rk} \left( x_r^{(k^{*},k)} - j y_r^{(k^{*},k)}
\right) \label{Eq:IL}
\end{eqnarray}
where $z_{k^*}$, $a_{rk}$, $x_r^{(k^{*},k)}$, and
$y_r^{(k^{*},k)}$ are defined in Section~II.

Equivalently, (\ref{Eq:IL}) can be rewritten as
\begin{eqnarray}
I\left(\varphi_k\left| {\cal L}_{l}^{k^{*}} \right.\right)
\!\!&=&\!\! \sqrt{\frac{\sigma_w^2 \gamma}{L}}  z_{k^*} \!\!
\sum_{r \in {\cal L}_{l}^{k^{*}}} \left(
{x^\prime}_{r}^{(k^{*},k)} - j
{y^\prime}_{r}^{(k^{*},k)} \right) \nonumber \\
\!\!&=&\!\! z_{k^*} \left( \sqrt{\frac{\sigma_w^2 \gamma}{L}}
\!\!  \sum_{r \in {\cal L}_{l}^{k^{*}}} {x^\prime}_{r}^{(k^{*},k)}
- j \sqrt{\frac{\sigma_w^2 \gamma}{L}}  \!\! \sum_{r \in {\cal
L}_{l}^{k^{*}}} {y^\prime}_{r}^{(k^{*},k)} \right) \label{yIaa}
\end{eqnarray}
where ${x^\prime}_{r}^{(k^{*},k)} \triangleq a_{rk}
x_r^{(k^{*},k)}$ and ${y^\prime}_{r}^{(k^{*},k)} \triangleq a_{rk}
y_r^{(k^{*},k)}$ with mean and variance given as
\begin{eqnarray}
m_1 \!\!&=&\!\! m_{a_{rk}} m_{u} = 0,\\
\sigma_1^2 \!\!&=&\!\! \left( \sigma_u^2 + m_u^2 \right) \left(
\sigma_{a_{rk}}^2 + m_{a_{rk}}^2 \right) - m_{a_{rk}}^2 m_u^2 =
\sigma_u^2 \sigma_{a_{rk}}^2.
\end{eqnarray}
Let us introduce the notations $X_{l}^{(k^{*},k)} \triangleq
\sqrt{\sigma_w^2 \gamma/L} \sum_{r \in {\cal L}_{l}^{k^{*}}}
{x^\prime}_{r}^{(k^{*},k)}$ and $Y_{l}^{(k^{*},k)} \triangleq
\sqrt{\sigma_w^2 \gamma / L}$  $\times \sum_{r \in {\cal
L}_{l}^{k^{*}}} {y^\prime}_{r}^{(k^{*},k)}$. Then
$X_{l}^{(k^{*},k)}$ and $Y_{l}^{(k^{*},k)}$ can be approximated by
Gaussian random variables \cite{Ochiai2005,Lo1964} with mean $ m_X
= m_Y = m_1 = 0$ and variance $\sigma_X^2 = \sigma_Y^2 = \gamma
\sigma_w^2 \sigma_1^2$. Thus, \eqref{yIaa} can be finally
rewritten as
\begin{eqnarray}
I\left(\varphi_k\left| {\cal L}_{l}^{k^{*}} \right.\right) =
z_{k^*} \left( X_{l}^{(k^{*},k)} - j Y_{l}^{(k^{*},k)} \right).
\label{yIbb}
\end{eqnarray}

Using \eqref{yIbb} and the fact that $| z_{k^*} |^2 = 1$, the
received interference power at the unintended BS/AP $d_{k}$ from
the candidate set of nodes ${\cal L}_{l}^{k^{*}}$
can be expressed as
\begin{equation}
\left|I\left(\varphi_k\left| {\cal L}_{l}^{k^{*}} \right.\right)
\right|^2 = \left( {X_{l}^{(k^{*},k)}}\right)^2 + \left(
{Y_{l}^{(k^{*},k)}}\right)^2 .
\end{equation}

The probability that the candidate
set of nodes ${\cal L}_{l}^{k^{*}}$ is approved to join the set of
collaborative nodes ${\cal N}^{k^{*}}$, i.e, the probability that
the INR $\eta$ from ${\cal L}_{l}^{k^{*}}$ at the
unintended BS/AP $d_k$ is lower than the threshold value
$\eta_{\text{thr}}$, can be then found as
\begin{eqnarray} \label{probunintinterf}
{\bf Pr} \left( \eta < \eta_{\text{thr}} \right) \!\!&=&\!\! {\bf
Pr} \left( \frac{\left| I(\varphi_k\left| {\cal L}_{l}^{k^{*}}
\right.) \right|^2 }{ \sigma_w^2} < \eta_{\text{thr}} \right) =
{\bf Pr} \left( \frac{\left( {X_{l}^{(k^{*},k)}}\right)^2 + \left(
{Y_{l}^{(k^{*},k)}}\right)^2
}{\sigma_w^2} < \eta_{\text{thr}} \right)  \nonumber \\ 
\!\!&=&\!\! 1 - \exp \left( - \frac{ \eta_{\text{thr}}
\sigma_w^2}{2 \sigma_X^2} \right) \triangleq p^\prime
\end{eqnarray}
where the INR $\eta = \left. \left\{\left(
{X_{l}^{(k^{*},k)}}\right)^2 + \left( {Y_{l}^{(k^{*},k)}}\right)^2
\right\} \right/\sigma_w^2$ is exponentially distributed random
variable with the probability density function (pdf)
\begin{eqnarray}
f\left(\eta \ |  \ \frac{\sigma_w^2}{2 \sigma_X^2}\right) =
\left\{\begin{matrix} \frac{\sigma_w^2}{2 \sigma_X^2}
\exp\left\{-\frac{\sigma_w^2 \eta}{2 \sigma_X^2}\right\}, &\; \eta
\ge 0 \\ 0, &\; \eta < 0 \end{matrix}\right. .
\end{eqnarray}

If $D$ unintended BSs/APs are present in the neighborhood of the
set of candidate collaborative nodes ${\cal L}_{l}^{k^{*}}$, the
probability that the INR from ${\cal L}_{l}^{k^{*}}$ at any one of
these unintended BSs/APs is lower than the threshold value
$\eta_{\text{thr}}$ is given by \eqref{probunintinterf}.
Therefore, the probability that ${\cal L}_{l}^{k^{*}}$ is approved
by all BSs/APs is the product of the probabilities that ${\cal
L}_{l}^{k^{*}}$ is approved by each of the unintended BSs/APs,
that is,
\begin{eqnarray} \label{FinProb}
p = \left( 1 - \exp \left( - \frac{ \eta_{\text{thr}}
\sigma_w^2}{2 \sigma_X^2} \right) \right)^{D}.
\end{eqnarray}
It can be seen from \eqref{FinProb} that $p$ decreases if the
threshold $\eta_{\text{thr}}$ decreases or the number $D$ of
neighboring BSs/APs increases.

Using \eqref{FinProb}, a closed-form expression for the average
number of trials required by the node selection algorithm can be
derived. Note that the actual number of trials is a random
variable which we will denote as $T$. In order to construct the
set of collaborative nodes ${\cal N}^{k^{*}}$, $T_0 = N/L$ trials
must be successful among all $T$ trials. Since the candidate nodes
are selected randomly at each trial of the node selection
algorithm, the algorithm itself can be viewed as a sequence of
Bernoulli trials. Since $T_0$ of these Bernoulli trials must be
successful in order to construct ${\cal N}^{k^{*}}\!\!\!$, the
probability distribution of $T$ in a sequence of Bernoulli trials
is, in fact, negative binomial distribution, that is,
\begin{eqnarray} \label{Binom}
{\bf Pr} (T=t) = \binom{t-1} {T_0 - 1} p^{T_0} \left( 1 - p
\right)^{t - T_0}.
\end{eqnarray}

Using \eqref{Binom}, the average number of trials for the proposed
node selection algorithm can be obtained as (see the details of
the derivation in Appendix B)
\begin{eqnarray} \label{NumTrials}
E \left\{ T\right \}  = \frac{T_0}{p} = \frac{N}{L \cdot p}.
\label{Eq:AvNoItr}
\end{eqnarray}
It can be seen from (\ref{NumTrials}) that the average number of
trials is proportional to the size of the set of collaborative
nodes ${\cal N}^{k^{*}}\!\!\!$, but it is inverse proportional to the
size of the candidate set of nodes ${\cal L}_{l}^{k^{*}}$ and to
the probability that the set ${\cal L}_{l}^{k^{*}}$ is approved to
join the set ${\cal N}^{k^{*}}$. Therefore, less number of trials
is required in average for the proposed node selection algorithm
if $L$ is chosen to be large or $N$ is small. Moreover, if the
probability $p$, which, in turns, depends on the threshold value
$\eta_{\text{thr}}$ of the INR allowed at the
unintended BSs/APs from the set ${\cal L}_{l}^{k^{*}}$ (see \eqref{FinProb}), is large,
then less number of trials is required.

\subsection{The Complementary Cumulative Distribution Function
of the Interference} Let us assume that $K (\le D)$ collaborative
sets are active and target different destination from $d_{k^{*}}$,
these sets are ${\cal N}^k$, $k \ne k^{*}$, and their union is
denoted hereafter as $\bigcup {\cal N}^{k \ne k^{*}}$. It can be
seen from (\ref{Eq:y}) that the total interference collected at
the destination $d_{k^{*}}$ from these $K$ collaborative sets is
\begin{eqnarray} \label{intyI}
I \left( \varphi_{k^{*}} \, \left| \, \bigcup {\cal N}^{k \ne
k^{*}} \right. \right) = \sqrt{\frac{\sigma_w^2 \gamma}{N}}
\sum_{k \ne k^{*}} z_k \sum_{r \in {\cal N}^k} a_{{k^{*}}r} \left(
x_r^{(k^{*},k)} - j y_r^{(k^{*},k)} \right)
\end{eqnarray}
where the power per one collaborative sensor node is $P =
\sigma_w^2 \gamma / N$ because the SNR at the intended BS/AP must
be $10 \log_{10}(\gamma)$~dB. Using the fact that ${\cal N}^k =
\bigcup {\cal L}_{l}^{k}$ and multiplying and dividing the right
hand side of \eqref{intyI} by $\sqrt{L}$, the total interference
at $d_{k^{*}}$ can be expressed as
\begin{eqnarray} \label{yIfinal}
I \left( \varphi_{k^{*}} \, \left| \, \bigcup {\cal N}^{k \ne
k^{*}} \right. \right) \!\!&=&\!\!  \sqrt{\frac{L}{N}} \sum_{k \ne
k^{*}} z_k \sum_{l = 1}^{N/L} \sum_{r \in {\cal L}_{l}^{k}}
\sqrt{\frac{\sigma_w^2 \gamma}{L}} \left(
{x^\prime}_{r}^{(k^{*},k)} -
j {y^\prime}_{r}^{(k^{*},k)} \right) \nonumber \\
\!\!&=&\!\! \sqrt{\frac{L}{N}} \sum_{k \ne k^{*}} z_k
\sum_{l=1}^{N/L} \left( \tilde{X}_{l}^{(k^{*},k)}  - j \
\tilde{Y}_{l}^{(k^{*},k)} \right)
\end{eqnarray}
where $\tilde{X}_{l}^{(k^{*},k)}$ and $\tilde{Y}_{l}^{(k^{*},k)}$
are zero mean truncated Gaussian distributed random variables
corresponding to $X_{l}^{(k^{*},k)}$ and $Y_{l}^{(k^{*},k)}$ of
\eqref{yIbb} for only the approved candidate subsets. It can be
shown that the marginal conditional probability density function
of $\tilde{U}_{l}^{(k^{*},k)} \in \left\{
\tilde{X}_{l}^{(k^{*},k)},\tilde{Y}_{l}^{(k^{*},k)}\right\}$ is
\cite{Helstrom}
\begin{eqnarray}
\!\!& &\!\! f \left( \left. \tilde{U}_{l}^{(k^{*},k)} \, \right|
\, \eta \le \ \eta_{\text{thr}} \right) \!=\! f \left(
\tilde{U}_{l}^{(k^{*},k)} \, \left| \, \left(
\tilde{X}_{l}^{(k^{*},k)} \right)^2 + \left(
\tilde{Y}_{l}^{(k^{*},k)} \right)^2 \le \sigma_w^2
\eta_{\text{thr}} \right. \right) \nonumber \\
\!\!& & \qquad  =\! \frac{1}{\sqrt{2 \pi \sigma_X^2}} \left[ 1 -
\exp \left( -\frac{\sigma_w^2 \eta_{\text{thr}}}{2 \sigma_X^2}
\right) \right]^{-1} \left[ 1 \!-\! 2 \ \text{Q} \left( \! \frac{
\sqrt{\sigma_w^2  \eta_{\text{thr}} \!-\! \left(
\tilde{U}_{l}^{(k^{*},k)}\right)^2}}{\sigma_X} \right) \right] \!
\nonumber \\
\!\!& & \qquad \qquad \times \exp \!\!\left( \!-\frac{\left(
\tilde{U}_{l}^{(k^{*},k)}\right)^2}{2 \sigma_X^2} \right)\!\!, \;
|\tilde{U}_{l}^{(k^{*},k)}| \!\!\le\!\! \sqrt{\sigma_w^2 \
\eta_{\text{thr}}}
\end{eqnarray}
where $\text{Q} (x) = 1 / \sqrt{2 \pi} \int_x^\infty \exp (- u^2 /
2 ) \, du$ is the Q-function of the Gaussian distribution.

Using \eqref{yIfinal}, the total INR at $d_{k^*}$ can be
then expressed as
\begin{eqnarray} \label{Inferfforone}
\eta &=& \frac{1}{\sigma_w^2}  \left| I \left( \varphi_{k^{*}} \,
\left|  \, \bigcup {\cal N}^{k \ne k^{*}} \right.\right) \right|^2
=  \frac{L}{N \sigma_w^2} \sum_{k \ne k^{*}} \left( \left( \sum_{l
= 1}^{N/L} {\tilde{X}_{l}^{(k^{*},k)}} \right)^2 + \left( \sum_{l
= 1}^{N/L}
{\tilde{Y}_{l}^{(k^{*},k)}} \right)^2 \right) \nonumber \\
&=& \sum_{k \ne k^{*}} \left( \left( \sqrt{\frac{L}{N \sigma_w^2}}
\sum_{l = 1}^{N/L} {\tilde{X}_{l}^{(k^{*},k)}} \right)^2 + \left(
\sqrt{\frac{L}{N \sigma_w^2}} \sum_{l = 1}^{N/L}
{\tilde{Y}_{l}^{(k^{*},k)}} \right)^2 \right) .
\end{eqnarray}

Based on the central limit theorem, both the real and imaginary
parts of the interference from each set of collaborative nodes,
i.e., $\sqrt{\frac{L}{N \sigma_w^2}} \sum_{l = 1}^{N/L}
\tilde{X}_{l}^{(k^{*},k)}$ and $\sqrt{\frac{L}{N \sigma_w^2}}
\sum_{l = 1}^{N/L} \tilde{Y}_{l}^{(k^{*},k)}$, are zero mean
Gaussian distributed random variables with variance for both given
as
\begin{eqnarray}
\sigma_I^2 \!\!&=&\!\!  E \left\{ \left( \sqrt{\frac{L}{N
\sigma_w^2}} \sum_{l = 1}^{N/L} \tilde{X}_{l}^{(k^{*},k)}\right)^2
\right\} = E \left\{ \left( \sqrt{\frac{L}{N \sigma_w^2}} \sum_{l
= 1}^{N/L} \tilde{Y}_{l}^{(k^{*},k)}\right)^2 \right\} \nonumber
\\
\!\!&=&\!\! \frac{\sigma_X^2 (1 - ( 1 + \beta ) e^{-\beta}
)}{\sigma_w^2(1 - e^{-\beta})}
\end{eqnarray}
where $\beta = \sigma_w^2 \eta_{\text{thr}}/2 \sigma_X^2$.

Since the real and imaginary parts of the interference from each
set of collaborative nodes are Gaussian distributed, the INR from
each set of collaborative nodes is exponentially distributed for a
given constant noise power $\sigma_w^2$. Therefore, the total INR
$\eta$ collected at $d_{k^*}$ from all collaborative sets is a sum
of exponentially distributed random variables (see
\eqref{Inferfforone}) and can be shown to be Erlang distributed,
that is,
\begin{eqnarray} \label{InerferenceDistribution}
f \left( \eta \, \left| \, K, \alpha \right. \right) = \frac{
\alpha^K {(\eta)}^{K-1} \exp \left( - \alpha \eta
\right)}{(K-1)!}, \quad \text{for} \; K > 0, \eta \ge 0, \; \alpha
= \frac{1}{2 \sigma_I^2}.
\end{eqnarray}

Finally, using \eqref{InerferenceDistribution}, the CCDF of the
INR, i.e., the INR distribution over the
the sidelobes of the CB beampattern with node selection, can be
expressed in closed-form as
\begin{eqnarray}
{\bf Pr} \left( \eta \ge  \eta_0 \right) = \sum_{k = 0}^{K-1}
\frac{{(\alpha \eta_0 )}^k e^{-\alpha \eta_0}}{k!} \label{EqnCCDF}
\end{eqnarray}
where $K (\le D)$ is the number of active sensor node clusters in the
neighborhood of the BS/AP ${d_{k^*}}$.

%
%

\section{Simulation results}
\label{sec:sim} In order to demonstrate the advantages of the
proposed node selection algorithm for the CB beampattern sidelobe
control and verify the accuracy of the analytical expressions, we
present also the following simulation results.

\subsection{Sample CB Beampattern}
In our first example, we study numerically the performance of the
proposed node selection algorithm for the CB beampattern sidelobe
control. Unless otherwise is specified, the nodes are assumed to
be uniformly distributed over a disk with radius $R = 2 \lambda$.
The total number of sensor nodes in the coverage area of the
transmitting source node is $M = 512$ and the desired number of
collaborative nodes to be selected is $N=256$. The power budget
per all $N$ collaborative nodes equals to 20~dB. The size of a
group of candidate sensor nodes $L$ is taken to be equal to $32$
and the INR threshold value at the unintended BSs/APs is set to
$\eta_{\text{thr}} = 10$~dB.

\begin{figure}[t]
\centering
\includegraphics[scale=0.8]{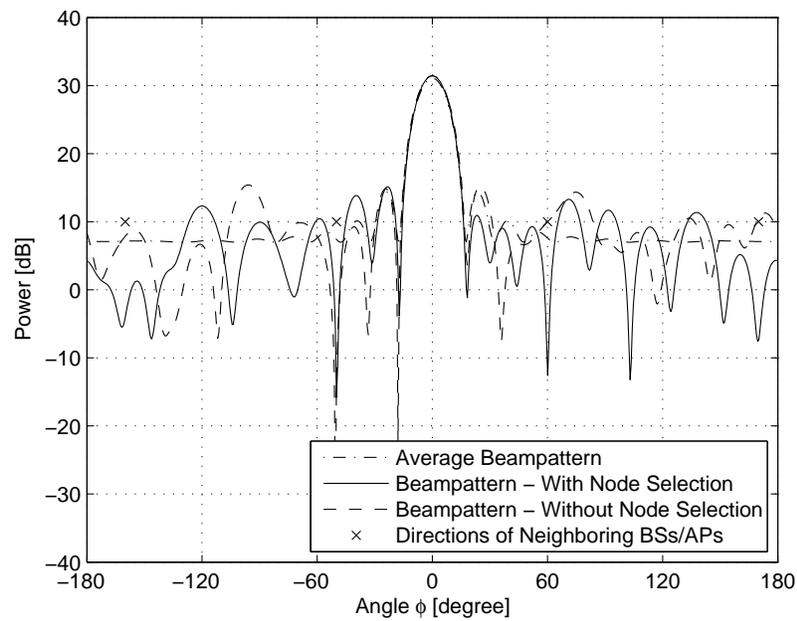}
\caption{Beampattern: The intended BS/AP is located at $\varphi_0 = 0^o$ and $4$ neighboring BSs/APs at directions $\varphi_1=-160^o$, $\varphi_2=-50^o$,  $\varphi_2=60^o$, and $\varphi_3=170^o$.}
\label{Fig:Journal2_fig2}
\end{figure}

\begin{figure}[t]
\centering
\includegraphics[scale=0.8]{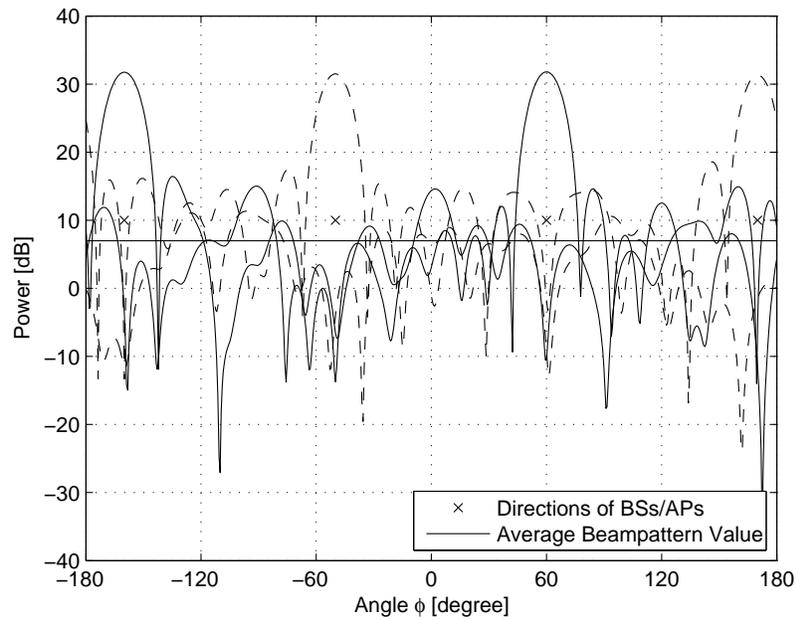}
\caption{Beampattern: $4$ simultaneous beampatterns with BSs/APs at directions $\varphi_1=-160^o$, $\varphi_2=-50^o$,  $\varphi_2=60^o$, and $\varphi_3=170^o$.}
\label{Fig:Journal2_fig3}
\end{figure}

\begin{figure}[t]
\centering
\includegraphics[scale=0.8]{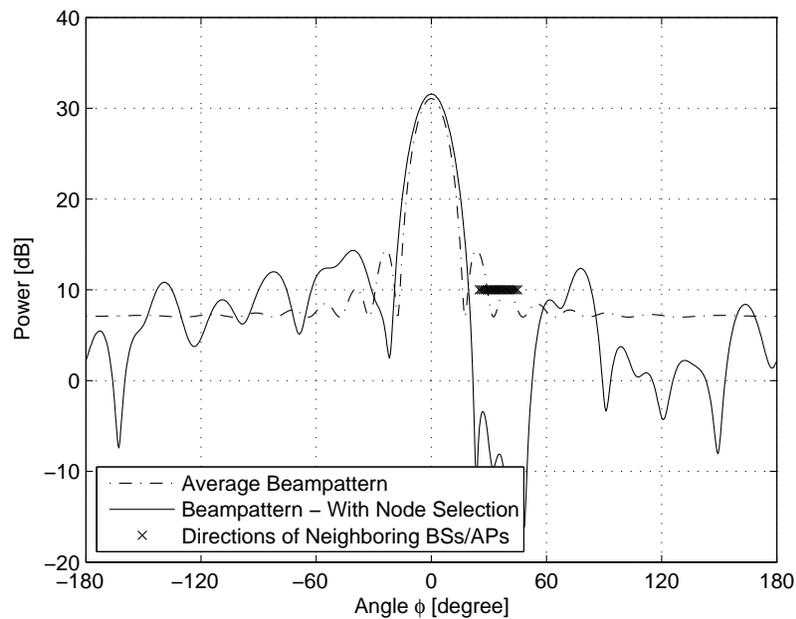}
\caption{Beampattern: The neighboring BSs/APs are located in the
range $\phi \in [ 25^o \, 45^o ]$ and $\eta_{\text{thr}} =
10$~dB.} \label{Fig:Journal2_fig2a_}
\end{figure}

\begin{figure}[t]
\centering
\includegraphics[scale=0.8]{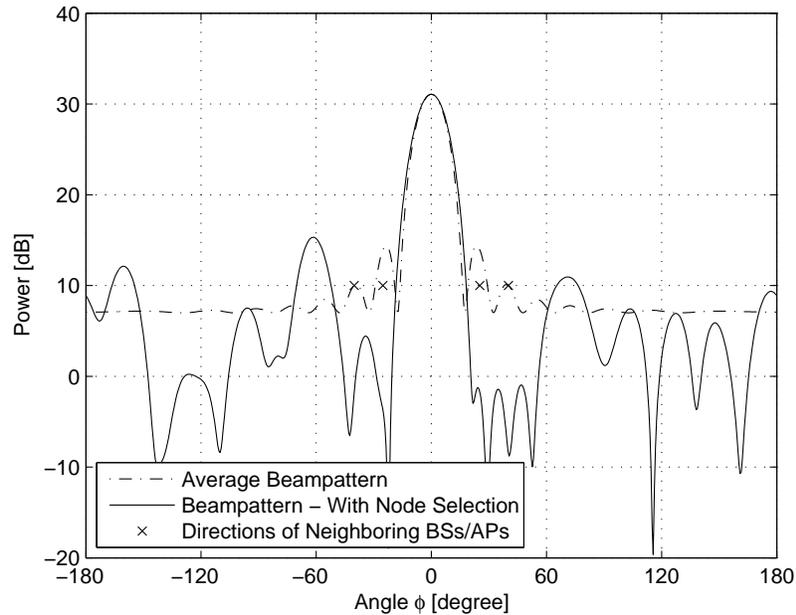}
\caption{Beampattern: The neighboring BSs/APs are at directions corresponding to the peaks of the average beampattern.}
\label{Fig:Journal2_fig2b}
\end{figure}

We report the results for the following four different cases.
\begin{itemize}
\item[{\it Case 1:}] The intended BS/AP is located at the direction
$\varphi_0 = 0^o$. There are $D = 4$ unintended neighboring
BSs/APs are present at the directions $\varphi_1=-160^o$,
$\varphi_2=-50^o$, $\varphi_3=60^o$, and $\varphi_4=170^o$.

In Fig.~\ref{Fig:Journal2_fig2}, we plot the sample beampattern
corresponding to the CB with node selection and compare it to the
sample beampattern corresponding to the CB without node selection
and the average beampattern. The directions to the unintended
BSs/APs are marked by symbol `$\times$'. It can be seen from the
figure that the CB with node selection archives the lowest
sidelobes in the directions of unintended BSs/APs, while the
sidelobes of the CB without node selection are uncontrolled and
high in the directions of unintended BSs/APs. Moreover, it can be
seen that due to the CB coherent processing gain the peak of the
mainlobe corresponds to 31~dB for all the beampatterns, while the
power budget per all $N$ collaborative nodes was set to 20~dB. The
latter can be also predicted by the theoretically computed gain
increase of $\log_{10} N$ (see \cite{Ochiai2005},
\cite{Ahmed2009a}).

\item[{\it Case 2:}] In this case, it is required that the data
from different sensor nodes have to be sent to all 4 BSs/APs
simultaneously. Fig.~\ref{Fig:Journal2_fig3} shows the CB
beampatterns for the corresponding 4 CB clusters of 256
collaborative nodes selected from 512 nodes available in each
cluster. Note that the sets of sensor nodes in all 4 CB clusters
are different from each other and do not overlap.

It can be seen from the figure that each beampattern has minimum
interference at the direction of the mainlobes of the other
beampatterns. Moreover, the mainlobes of the corresponding
beampatterns all have the required mainlobes with a peak value of
about 31~dB.

\item[{\it Case 3:}] In our third case, the neighboring BSs/APs are
assumed to be located in the range $\phi \in [25^o \, 45^o ]$
which is closed to the mainlobe direction of the intended BS/AP.
The INR threshold value is set to $\eta_{\text{thr}} = 10$~dB.

The beampattern of the CB with node selection and the average
beampattern are shown in Fig.~\ref{Fig:Journal2_fig2a_}. It can be
seen from the figure that the CB with node selection is able to
achieve a beampattern with sufficiently low sidelobes over the
whole range $\phi \in [25^o \, 45^o ]$. Note that this case
corresponds to the situation when the unintended AP is actually
another cluster of sensor nodes distributed over space, which,
therefore, cannot be viewed as a point in space.

\item[{\it Case 4:}] In the last case, we assume that $D=4$
neighboring untended Bs/APs are located at the angles
corresponding to the peaks of the average beampattern, while the
intended BS/AP is located at $\varphi_0 = 0^o$.

Fig.~\ref{Fig:Journal2_fig2b} shows the average beampattern and
the beampattern of the CB with node selection. Note that the peaks
of the average beampattern are located close to the mainlobe of
the average beampattern. Therefore, the locations of the
unintended BSs/APs are actually the worst locations in terms of
the corresponding average interference levels. As it can be seen
from the figure, using the node selection, we can achieve minimum
interference levels at the directions of unintended BSs/APs even
in this case.
\end{itemize}
Summarizing, it can be concluded based on all these cases that
there is an significant improvement achieved by using the node
selection algorithm in reducing the sidelobe levels at the
directions of unintended BSs/APs.

\subsection{Effect of The Algorithm Parameters}
The two parameters in the node selection algorithm are the INR
threshold $\eta_{\text{thr}}$ and the size $L$ of the candidate
set of nodes ${\cal L}^{k^{*}}\!\!$.

In this example, it is assumed that the intended BS/AP is located
at $\varphi_0 = 0^o$ and there is one unintended neighboring BS/AP
at the direction $\varphi_1=65^o$. The noise power equals to
$\sigma_{w}^2 = 0.05$. The coverage area of the source node has a
radius $R = 5 \lambda$. The INR threshold value changes in the
range $\eta_{\text{thr}} = [-15 \, 10 ]$~dB. The parameters of the
Gaussin distribution corresponding to the lognormal distribution
of the channel coefficients are $m = 0$ and $\sigma^2  = 0.2$.
Monte Carlo simulations are carried over using $1000$ runs to
obtain average results.

\begin{figure}[t]
\centering
\includegraphics[scale=0.8]{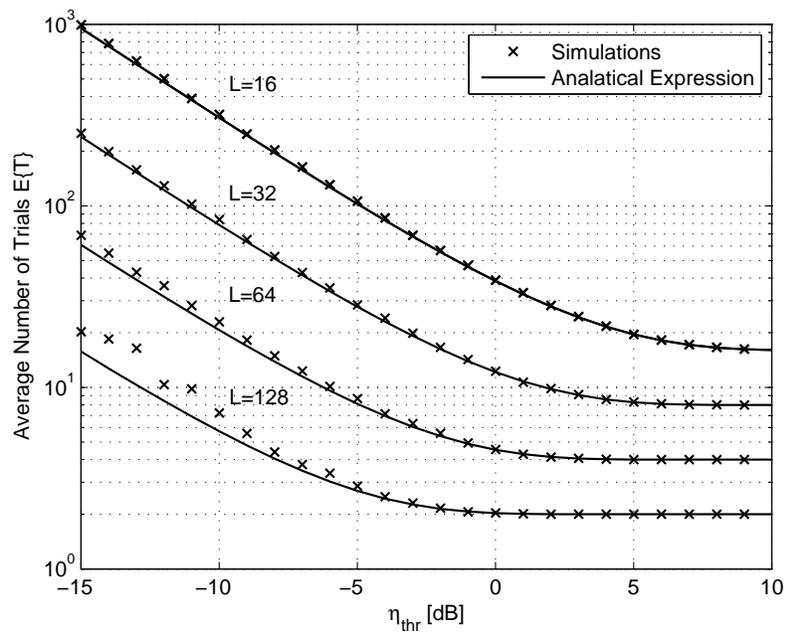}
\caption{Average number of trials $E\left\{T\right\}$ versus
threshold $\eta_{\text{thr}}$: $M = 512$, $N = 256$,
$\varphi_0=0^o$, and $\varphi_1=65^o$.}
\label{Fig:Journal2_fig4_1}
\end{figure}

Fig.~\ref{Fig:Journal2_fig4_1} demonstrates the effect of the
threshold $\eta_{\text{thr}}$ on the average number of trials
required to select the set of collaborative nodes ${\cal N}^k$
using the sets of candidate nodes of different sizes $L \in \{ 16,
32, 64, 128 \} $. It can be seen from the figure that the curves
obtained using the closed-form expression (\ref{Eq:AvNoItr}) for
the number of trials are in good agreement with the simulation
results. It can also be seen that by decreasing the threshold
$\eta_{\text{thr}}$, the number of trials increases. Moreover, the
number of trials can be controlled using $L$. Indeed, as $L$
increases, the number of trials decreases. It is important to note
that because of the normalization factor in (\ref{Eq:IL}), the
consumed power at each trial is the same for different values of
$L$ and the total consumed power in the selection process is
proportional to the number of trials.

\begin{figure}[t]
\centering
\includegraphics[scale=0.8]{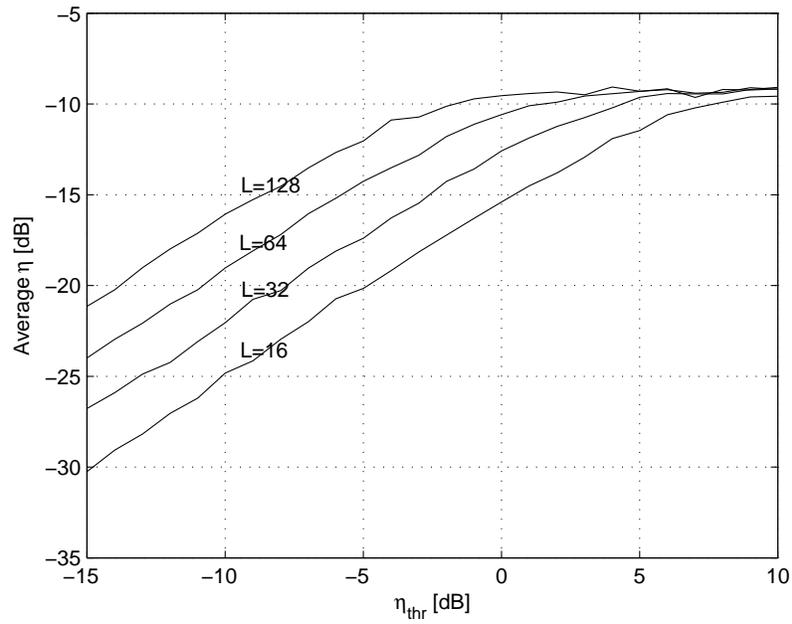}
\caption{Average INR versus threshold $\eta_{\text{thr}}$: $M =
512$, $N = 256$, $\varphi_0=0^o$, and $\varphi_1=65^o$.}
\label{Fig:Journal2_fig4_2}
\end{figure}

Fig.~\ref{Fig:Journal2_fig4_2} shows the average interference
level of the CB beampattern with node selection versus threshold
$\eta_{\text{thr}}$ for different values of $L \in \{ 16, 32, 64,
128 \} $. It can be seen from the figure that the average
interference level is proportional to the threshold
$\eta_{\text{thr}}$. Comparing
Figs.~\ref{Fig:Journal2_fig4_1}~and~\ref{Fig:Journal2_fig4_2} to
each other, we can observe a tradeoff between the average number
of trials required for node selection and the achieved average
interference (sidelobe) level. It can be seen that lower
interference level can be achieved by using smaller values of $L$
at the expense of larger number of trials.

In addition, it is worth noting that there is no limitations in
the node selection algorithm on selecting the value of $L$ as long
as $L \le N$. However, the threshold $\eta_{\text{thr}}$ affects
the average interference level and, therefore, depends on the
sensitivity of the front--end receiver of the BS/AP.

\subsection{Effect of Number of Neighboring BSs/APs}
In this example, we study the effect of number of neighboring
BSs/APs $D$ to the performance of the node selection algorithm.
Note that for simplicity we have always assumed that
$\eta_{\text{thr}}$ is the same for all BSs/APs. Such set up does
not restrict the generality of the proposed node selection
algorithm since the selection is performed based on the {\it
accept/regect} bit from the corresponding BS/AP, while
$\eta_{\text{thr}}$ is used only in the BS/AP. Therefore, it is
straightforward to use different threshold values at different
BSs/APs, and no changes to the node selection algorithm have to be
done.

\begin{figure}[t]
\centering
\includegraphics[scale=0.8]{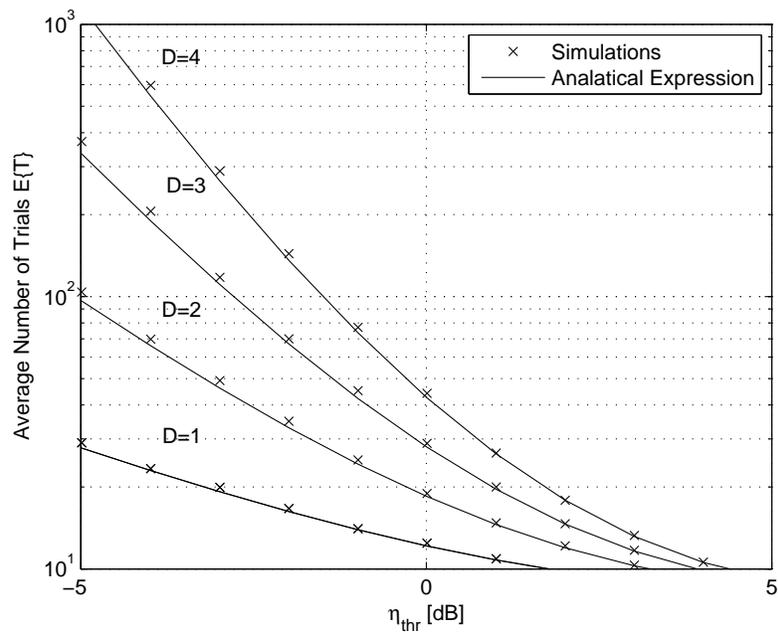}
\caption{Average \ number \ of \ trials \ $E\left\{T\right\}$ \ versus \ threshold \ $\eta_{\text{thr}}$ \ for \ different \ values \ of \ $D$.}
\label{Fig:Journal2_fig5_1}
\end{figure}

In Fig.~\ref{Fig:Journal2_fig5_1}, the average number of trials is
plotted versus the threshold $\eta_{\text{thr}}$ for different
values of $D \in \{ 1, 2, 3, 4\}$. It can be seen from this figure
that as $D$ increases, the average number of trials of the node
selection algorithm increases exponentially if the required INR
threshold $\eta_{\text{thr}}$ is low. Finally, it can be observed
that the analytical and simulation results are in a good agreement
with each other.

\subsection{The CCDF of the Beampattern Level}
In our last example, we investigate the CCDF of the beampattern
level.

\begin{figure}[t]
\centering
\includegraphics[scale=0.8]{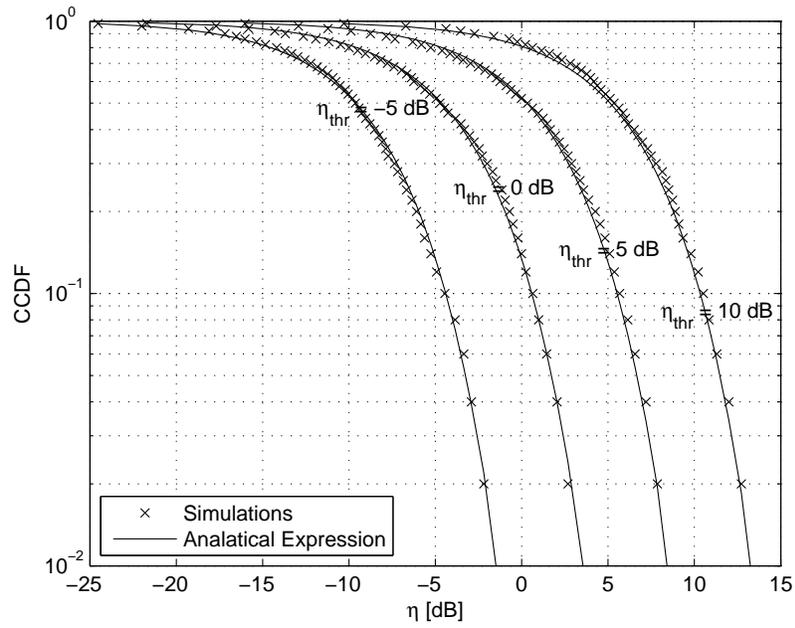}
\caption{The CCDF of the INR for different values of the threshold
$\eta_{\text{thr}}$: $M = 512$, $N = 256$, $L = 32$,
$\varphi_0=0^o$, and $\varphi_1=65^o$.} \label{Fig:Journal2_fig6}
\end{figure}

\begin{figure}[t]
\centering
\includegraphics[scale=0.8]{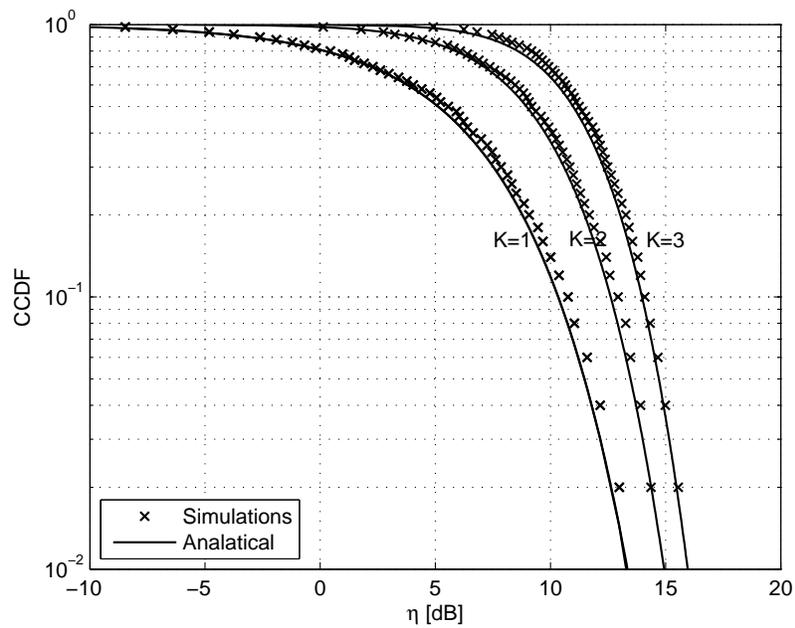}
\caption{The CCDF of the INR for different values of $K$: $M =
512$, $N = 256$, $L = 32$, $\varphi_0=0^o$, and $\eta_{\text{thr}}
= 10$~dB.} \label{Fig:Journal2_fig7}
\end{figure}

Fig.~\ref{Fig:Journal2_fig6} depicts the probability that the
interference exceeds certain level, i.e., it shows the CCDF of
interference for different values of the INR threshold
$\eta_{\text{thr}} \in \{ -5, 0, 5, 10 \}$. In addition,
Fig.~\ref{Fig:Journal2_fig7} illustrates the CCDF of the
interference for different numbers of active collaborative sets $K
\in \{ 1, 2, 3 \}$. It can be seen from
Fig.~\ref{Fig:Journal2_fig6} that the CCDF of the interference
increases as $\eta_{\text{thr}}$ decreases. Moreover, as can be
observed from Fig.~\ref{Fig:Journal2_fig7}, the CCDF of the
interference increases if $K$ increases. The latter fact agrees
with the intuition that for larger number of collaborative sets
transmitting simultaneously, the overall received interference by
all BSs/APs must be higher. The simulation results in both figures
perfectly agree with our analytical results as well.

\section{CONCLUSIONS}
Node selection is introduced for the CB sidelobe control in the
context WSNs. A low-overhead and efficient node selection
algorithm is developed and analyzed. In particular, the
expressions for the average number of trials required for the
proposed node selection algorithm and the CCDF of the beampattern
sidelobe level of the CB with node selection are derived. The
effect of the number of nodes selected at one trial to the
algorithm performance is also investigated. It is shown that
increasing the number of nodes selected at one trial reduces the
number of trials at the expense of higher sidelobe levels. It is
also shown that the CCDF of the beampattern level depends on the
interference threshold value. {F}rom both the analytical and
simulation results, we have seen that CB with node selection has
perfect interference suppression capabilities as compared to the
CB without node selection for which the beampattern sidelobes are
uncontrolled and can cause significant interference to unintended
BSs/APs.

\section*{Appendix A: Derivation of the Mean and Variance of $x_r^{(k^{*},k)}$ and
$y_r^{(k^{*},k)} $}

First, we find the probability distribution of $u \in \left\{ x_r^{(k^{*},k)},
y_r^{(k^{*},k)} \right\}$ and then find the corresponding mean and variance.
Assuming that the angles $\theta_r^{k}$ and $\theta_r^{k^{*}}$ are uniform
distributed in the interval $[-\pi,\pi]$, i.e., $\theta \sim
{\cal U}[-\pi,\pi]$, it can be found that the
difference $\Delta = \theta_r^{k} - \theta_r^{k^{*}}$ has the
following distribution
\begin{eqnarray} \label{Function}
f(\Delta) = \left\{
\begin{tabular}{ll}
$\frac{2 \pi - \Delta}{4\pi^2}$, & {$0 \le \Delta \le 2 \pi$;} \\
$\frac{2 \pi + \Delta}{4\pi^2}$, & {$-2 \pi \le \Delta \le 0$.}
\end{tabular}
\right.
\end{eqnarray}

Using the equality $u = {\cal R} \left\{ e^{j\Delta} \right\} =
\cos \left( \Delta \right)$, we can find the roots of the equation
$u = \cos \left( \Delta \right)$ in the interval $[-2\pi, 2\pi]$
as
\begin{eqnarray} \label{Xroots}
\Delta_1 &=& \cos^{-1} (u) \nonumber \\
\Delta_2 &=& \pi + \cos^{-1} (u) .
\end{eqnarray}

Then the distribution of $u$ can be found using the well known
expression \cite{Papoulis}
\begin{eqnarray} \label{Papdest}
f (u) = \left. \frac{f (\Delta)}{ |f^\prime (\Delta) | }
\right|_{\Delta_1,\Delta_2}
\end{eqnarray}
where $f^\prime (\Delta)$ is the first derivative of $f (\Delta)$.

Substituting \eqref{Function} and \eqref{Xroots} into
\eqref{Papdest}, and using the facts that $|u'| = | - \sin
(\Delta) |$ and $\sin(\cos^{-1}(u)) = \sqrt{1-u^2}$, the
distribution of $u$ can be found as
\begin{eqnarray}
f(u) \!\!&=&\!\! \left. \frac{f (\Delta)}{\sin (\Delta)}
\right|_{\Delta_1, \Delta_2} \nonumber \\
\!\!&=&\!\! \frac{2 \pi - \cos^{-1}(u)}{2 \pi^2 \sin
(\cos^{-1}(u))} + \frac{\pi - (\pi +\cos^{-1} (u))}
{2 \pi^2 \sin(\pi+\cos^{-1}(u))} \nonumber \\
\!\!&=&\!\! \frac{2 \pi}{2 \pi^2 \sqrt{1-u^2}} = \frac{1}{\pi
\sqrt{1 - u^2}}.
\end{eqnarray}

The mean can be then easily found as
\begin{eqnarray}
m_u = E \left\{ u \right \} = \int_{-1}^{1} u f(u) du  =
\int_{-1}^{1} \frac{u}{\pi \sqrt{1 - u^2}} du = \left.
-\sqrt{1-u^2}\right|_{-1}^{1}=0.
\end{eqnarray}

And the variance can be found as
\begin{eqnarray}
\sigma_u^2 = E \left\{ u^2 \right\} = \int_{-1}^{1} u^2 f(u) du =
\int_{-1}^{1}  \frac{u^2}{\pi \sqrt{1 - u^2}} du = \left.
\frac{\sin^{-1}(x)}{2\pi} \right|_{-1}^{1} - \left.
\frac{u\sqrt{1-u^2}}{2\pi}\right|_{-1}^{1}= 0.5.
\end{eqnarray}
Similar result can be also derived for $u \triangleq {\cal I}
\left\{ e^{j\Delta} \right\}=\sin \left( \Delta \right)$ in the
same way.

\section*{Appendix B: Derivation of \eqref{NumTrials} }

Consider an infinite sequence of independent Bernoulli trials with
probability of success $p$. Let $Z_1$ denotes the number of trials
before the first successful trial. Then $Z_1$ is geometric
distributed random variable $Z_1 \sim \text{Geom}(p)$, that is,
\begin{eqnarray} \label{Geom}
{\bf Pr} (Z_1 = k) = (1 - p)^{k-1} p, \quad k = 1, 2, \cdots,
\infty .
\end{eqnarray}

The corresponding moment generating function (MGF) for
\eqref{Geom} is
\begin{eqnarray}
M_{Z_1} (t) \!\!&\triangleq&\!\! E \{ e^{t Z_1} \} =
\sum_{k=1}^{\infty} e^{t k} (1-p)^{k-1} p \nonumber \\
\!\!&=&\!\! p e^{t} \sum_{l=0}^{\infty} \left( (1-p) e^{t}
\right)^{l} = \frac{p e^{t}}{1 - (1 - p) e^{t}}.
\end{eqnarray}

Therefore, the average value of $Z_1$ can be found as
\begin{eqnarray}
E \{Z_1 \} \!\!&=&\!\! \left. \frac{d}{dt} M_{Z_1} (t)
\right|_{t=0} = \left. \frac{(1 - (1 - p) e^{t}) (p e^{t}) - ((p -
1) e^{t}) (p e^{t})}{(1 - (1 - p) e^{t})^2} \right|_{t=0} \nonumber \\
\!\!&=&\!\! \left. \frac{p e^{t}}{(1 - (1 - p) e^{t})^2}
\right|_{t=0} = \frac{p}{p^2} = \frac{1}{p}.
\end{eqnarray}

Similarly, we can find the average number of trials between the
first and second successful trials $E \{Z_2 \}$, second and third
successful trials $E \{Z_3 \}$, and so on until $T_0 \triangleq
N/L$ successful trial. Since $Z_1, Z_2, \cdots , Z_{T_0}$ are
independent identically geometric distributed random variables,
i.e., $Z_i \sim \text{Geom} (p)$, $\sum_{i=1}^{T_0} Z_i$ is
negative geometric distributed, i.e., $\sum_{i=1}^{T_0} Z_i \sim
\text{NegBin}(T_0, p)$, with average
\begin{eqnarray}
E \left\{ \sum_{i=1}^{T_0} Z_i \right\} = T_0 E \{ Z_i \} =
\frac{T_0}{p} = \frac{N}{L \cdot p}
\end{eqnarray}
where the fact that $Z_1, Z_2, \cdots , Z_{T_0}$ are independent
identically distributed is used again.

\newpage

\end{document}